\documentclass[usenatbib]{mnras}
\usepackage{graphicx}
\usepackage{amssymb, amsmath}
\usepackage{times}
\usepackage{color}
\usepackage{multirow}
\usepackage{lscape}
\usepackage{varwidth}
\usepackage[section]{placeins}
\usepackage{epstopdf}
\usepackage{epsf}
\usepackage{float}
\usepackage{soul}

\newcommand{\cmnt}[1]{}

\newcommand{\Chris}[1]{{\bfseries{\textcolor{red}{#1}}}}

\begin{document}
\setlength{\parskip}{0pt}

\title[Post-reionization 21 cm streaming velocity BAO]{Streaming Velocity Effects on the Post-reionization 21 cm Baryon Acoustic Oscillation Signal}

\author[Long et~al.]{Heyang Long$^{1,2}$\thanks{E-mail: long.1697@osu.edu}, Jahmour J. Givans$^{3,4}$, and
Christopher M. Hirata$^{1,2,5}$\\
$^1$Department of Physics, The Ohio State University, 191 West Woodruff Avenue, Columbus, Ohio 43210, USA\\
$^2$Center for Cosmology and AstroParticle Physics (CCAPP), The Ohio State University, 191 West Woodruff Avenue, Columbus, Ohio 43210, USA\\
$^3$Department of Astrophysical Sciences, Princeton University, 4 Ivy Lane, Princeton, New Jersey 08540, USA\\
$^4$Center for Computational Astrophysics, Flatiron Institute, 162 5th Ave, New York, New York 10010, USA\\
$^5$Department of Astronomy, The Ohio State University, 140 West 18th Avenue, Columbus, Ohio 43210, USA
}

\date{2021 July 15} 
\pagerange{\pageref{firstpage}--\pageref{lastpage}} \pubyear{2021}
\maketitle
\label{firstpage}

\begin{abstract}
The relative velocity between baryons and dark matter in the early Universe can suppress the formation of small-scale baryonic structure and leave an imprint on the baryon acoustic oscillation (BAO) scale at low redshifts after reionization. This "streaming velocity" affects the post-reionization gas distribution by directly reducing the abundance of pre-existing mini-halos ($\lesssim 10^7 M_{\bigodot}$) that could be destroyed by reionization and indirectly modulating reionization history via photoionization within these mini-halos. In this work, we investigate the effect of streaming velocity on the BAO feature in H\,{\sc i} 21 cm intensity mapping after reionization, with a focus on redshifts $3.5\lesssim z\lesssim5.5$. We build a spatially modulated halo model that includes the dependence of the filtering mass on the local reionization redshift and thermal history of the intergalactic gas. In our fiducial model, we find isotropic streaming velocity bias coefficients $b_v$ ranging from -0.0043 at $z=3.5$ to -0.0273 at $z=5.5$, which indicates that the BAO scale is stretched (i.e., the peaks shift to lower $k$). In particular, streaming velocity shifts the transverse BAO scale between 0.121\% ($z=3.5$) and 0.35\% ($z=5.5$) and shifts the radial BAO scale between 0.167\% ($z=3.5$) and 0.505\% ($z=5.5$). These shifts exceed the projected error bars from the more ambitious proposed hemispherical-scale surveys in H\,{\sc i} (0.13\% at $1\sigma$ per $\Delta z = 0.5$ bin). 

\end{abstract}


\section{Introduction}
\label{sec:intro}
One of the main goals of cosmology is to understand the composition of the Universe and how it has evolved over time. General relativity (GR) relates the global properties of the Universe -- its mean density and pressure -- to the geometry of spacetime. Measurements of the geometry using the luminosity distance to Type Ia supernovae \citep{1998AJ....116.1009R, 1999ApJ...517..565P} revealed that the expansion of the Universe is accelerating; if interpreted within the framework of GR, this means that the bulk of the cosmic energy density is in the form of “dark energy”, which has negative pressure. This discovery has motivated a range of observational programs to precisely measure the expansion history of the Universe. These programs aim to measure whether the dark energy density is constant with time (a cosmological constant), or if it is varying, or if there might have been additional components to the cosmic energy budget at higher redshift.

One of the key methods of measuring cosmic geometry uses the baryon acoustic oscillations (BAO). These are acoustic oscillations in the optically thick photon-baryon plasma that filled the Universe before recombination that are seeded by the initial perturbations. At the time of recombination, the Universe becomes transparent; the baryons, no longer kinematically coupled to the photons, could gravitationally cluster to make large scale structures. The power spectrum of matter perturbations at low redshift contains oscillations as a function of wave number $k$ that are due to the phase of the acoustic oscillation at recombination \citep{1970Ap&SS...7....3S, 1970ApJ...162..815P}. These oscillations can be used as a standard ruler, whose length is set by early Universe physics \citep{1998ApJ...496..605E, 2002ASPC..280...35E} and can be calibrated using cosmic microwave background (CMB) observations. In a redshift survey, the use of the ``ruler'' is possible in both the transverse (measurement of distance) and radial (measurement of Hubble rate) directions \citep[e.g.,][]{2003ApJ...598..720S}.

The BAO feature in the distribution of matter can be computed robustly by solving the coupled Einstein, Boltzmann, and hydrodynamic equations of linear perturbation theory \citep[e.g.,][]{1995ApJ...455....7M} and using $N$-body simulations to follow the nonlinear evolution at low redshift (which leads only to modest changes in the standard ruler length; e.g., \citealt{2005Natur.435..629S}). However, the matter density field is not observable directly, particularly since 84 per cent of all matter in the Universe is dark matter \citep{2020A&A...641A...1P}. Instead we use visible tracers of the matter to measure BAOs. Most of the early measurements of BAOs were performed using massive, mostly red galaxies \citep[e.g.,][]{2005ApJ...633..560E,  2012MNRAS.427.2132P, 2012MNRAS.427.3435A, 2014MNRAS.439...83A, 2014MNRAS.441...24A, 2017MNRAS.464.1168R, 2017MNRAS.464.3409B, 2017MNRAS.470.2617A, 2020MNRAS.498.2492G, 2021MNRAS.500..736B}. Recent measurements have included star-forming galaxies \citep[e.g.,][]{2011MNRAS.415.2892B, 2014MNRAS.441.3524K, 2017MNRAS.464.4807H, 2021MNRAS.500.3254R, 2021MNRAS.501.5616D}, which have strong emission lines. Emission line galaxies are of particular interest for intermediate redshifts ($0.7\lesssim z\lesssim 2.5$) because the lines tend to be stronger than in the local Universe ($z\sim 0$), and the bright lines can be observed with much shorter exposures than would be required to measure the continuum of the galaxies (which is very faint due to the increasing luminosity distance). They will be targeted by ambitious new surveys such as DESI \citep{2016arXiv161100036D}, PFS \citep{2014PASJ...66R...1T}, {\slshape Euclid} \citep{2011arXiv1110.3193L}, and {\slshape Roman} \citep{2015arXiv150303757S}. One can also measure the BAO feature using neutral gas: at $z>1.9$, the Lyman-$\alpha$ forest is accessible from the ground; BAOs can be measured in the correlation function of the Lyman-$\alpha$ absorption \citep{2013A&A...552A..96B, 2013JCAP...04..026S, 2017A&A...603A..12B, 2019A&A...629A..85D} and in the correlation of this absorption with quasars \citep{2013JCAP...05..018F, 2017A&A...608A.130D, 2019A&A...629A..86B}.

Although there is an enormous volume potentially available for BAO studies at high redshifts, $2\lesssim z\lesssim 6$, upcoming galaxy surveys will only scratch the surface of the cosmological information available there. Individual galaxies become very faint, and their optical emission lines shift farther into the infrared (e.g., H$\alpha$ is beyond the red limit for both the {\slshape Euclid} and {\slshape Roman} space telescopes). The Lyman-$\alpha$ forest provides an alternative approach, and has given our current BAO constraints at $z\sim 2.4$, but it is sparsely sampled and as one increases the density of sightlines or probes higher redshift, one must go to fainter and fainter sources \citep{2011MNRAS.415.2257M}. H\,{\sc i} 21 cm emission observed in intensity mapping -- that is, in fluctuations in the diffuse background rather than individually detected galaxies -- has long been recognized as a powerful way to probe this range \citep{2008MNRAS.383.1195W, 2008PhRvL.100i1303C}. Measuring the 21 cm signal is observationally challenging due to the bright foregrounds and consequent need for exquisite control of instrumentation systematics (see, e.g., \citealt{1999A&A...345..380S} for an early discussion, and \citealt{2019MNRAS.483.2207M} for a recent discussion). But in the meter-wave radio band, it is possible to deploy enormous amounts of collecting area, and the digital signal processing required to calibrate many-element arrays and convert raw data into sky maps is advancing rapidly.

Current H\,{\sc i} intensity mapping efforts are focused on the lower redshifts where the foregrounds are fainter. These include CHIME \citep{2014SPIE.9145E..22B}, HIRAX \citep{2016SPIE.9906E..5XN}, TianLai \citep{2012IJMPS..12..256C}, FAST \citep{2011IJMPD..20..989N}, and BINGO\footnote{URL: http://www.bingotelescope.org/en/}. However, larger experiments probing the higher redshifts have been proposed: the ``Stage II'' Packed Ultrawide-band Mapping Array (PUMA) reference concept, for example, would be an interferometer composed of 32,000 dishes and probe half the sky out to $z\approx 6$ \citep{2018arXiv181009572C}. This would saturate most of the BAO information available in that hemisphere, reaching a statistical uncertainty in the BAO scale of 0.13\% per $\Delta z = 0.5$ bin at $3.5<z<5.5$.\footnote{This is calculated based on the effective number densities in \citet{2018arXiv181009572C} and the forecasting equations in \citet{2007ApJ...665...14S} with no reconstruction.}

This ambitious program will require both strong control of observational systematics, and an understanding of the astrophysical systematic errors in 21 cm BAO measurements. The BAO feature is famous for being more robust against astrophysical systematics than the broadband signal since complicated astrophysical processes are unlikely to produce a narrow feature in the correlation function at a specific scale. However, there is an important exception.
The same physics responsible for BAOs also gave baryons a supersonic streaming velocity relative to dark matter at decoupling, which has a feature at the same scale. If the tracer used for BAO analysis retains memory of the initial streaming velocity, then the BAO feature is distorted and shifted. This leads to an error in the Hubble parameter, $H(z)$, and angular diameter distance, $D_A(z)$, and hence in the inferred expansion history of the Universe. Previous work has explored the impact of streaming velocity in the power spectra of galaxies \citep{2010PhRvD..82h3520T,2010JCAP...11..007D,2011JCAP...07..018Y,2016PhRvL.116l1303B,2016PhRvD..94f3508S,2018MNRAS.474.2109S,2018ApJ...869...76A}, the Lyman-$\alpha$ forest \citep{2018MNRAS.474.2173H, 2020PhRvD.102b3515G}, reionization history \citep{2021ApJ...908...96P}, and the pre-reionization 21 cm field \citep{2019PhRvD.100f3538M, 2020ApJ...898..168C}.

This paper presents a first attempt to estimate the streaming velocity effect on post-reionzation 21 cm intensity mapping surveys. We focus our attention on the redshift range $3.5\lesssim z\lesssim 5.5$, i.e., after hydrogen reionization but before the bulk of He\,{\sc ii} reionization (and the associated complexities). We consider two major contributions to the streaming velocity bias $b_v$ (that is, the fractional change in H\,{\sc i} intensity in a region with the rms streaming velocity relative to a region with no streaming velocity). The first is the ``direct'' contribution, in which gas with a different streaming velocity ends up with a different temperature-density relation and different filtering scale after reionization. The second is the ``indirect'' contribution, in which the streaming velocity affects the clumping of the gas, and this (locally) changes the reionization history itself. We find the two effects to have the same order of magnitude, although we estimate the indirect effect to be larger. Our predicted angle-averaged BAO peak shifts range from $\Delta\alpha = -0.14\%$ at $z=3.5$ to $\Delta\alpha=-0.42\%$ at $z=5.5$, which would be significant for an experiment such as PUMA; but we caution that our estimates here represent only a first order-of-magnitude calculation of the BAO peak shift, and we identify possible future improvements.

This paper is organized as follows. In \S \ref{sec:formalism}, we define our formalism and biasing coefficients. In \S\ref{sec:methodology}, we lay out our program for estimating the various coefficients with a combination of simulations and analytic arguments. The simulations are presented in \S\ref{sec:methodology}, and the results in \S \ref{sec:results}. We conclude in \S \ref{sec:discussion}.
Some useful formulae for the filtering scale are given in Appendix~\ref{app:filtering}.

\section{Formalism and conventions}
\label{sec:formalism}

\subsection{Power Spectra}

In this paper, we focus on the power spectrum as the main 21 cm observable. (BAO measurements can be done in either correlation function space or power spectrum space, but for 21 cm both the theory and observations are native to the Fourier domain.) The observed quantity in 21 cm experiments is the differential brightness temperature $\Delta T_b$, which is defined as the observed brightness temperature $T_b$ relative to the CMB background temperature.  After reionization, the 21 cm signal comes mainly from neutral hydrogen in galaxies (which can be self-shielded) rather than the intergalactic medium. Assuming a high spin temperature $T_{\rm s}\gg T_{\rm CMB}$ typical of the neutral phase in galaxies, and neglecting self-absorption, the brightness temperature fluctuation (see, e.g., Eqs.~15--18 of \citealt{2006PhR...433..181F}) is proportional to the H\,{\sc i} density: 
\begin{equation}\label{eq:Tb}
    \Delta T_b(\bold{r},z)=\frac{3hc^3A_{10}(1+z)^2}{32\pi k_{\rm B}m_{\rm H}\nu_0^2H(z)} \frac{\rho_{\rm HI}(\bold{r},z)}{[\partial v_\parallel/\partial r_\parallel](1+z)/H(z)},
\end{equation}
where $\rho_{\rm HI}$ is the {\em comoving} density of H\,{\sc i}, $h$ is Planck's constant, $c$ is the speed of light, $k_{\rm B}$ is Boltzmann's constant, $A_{10}$ is the 21 cm Einstein cofficient, $m_{\rm H}$ is the mass of the hydrogen atom, $\nu_0$ is the 21 cm frequency, and $\partial v_\parallel/\partial r_\parallel$ is the line-of-sight velocity gradient. The second denominator in Eq.~(\ref{eq:Tb}) is unity in a homogeneous universe; it encodes the usual redshift space volume factor \citep{1987MNRAS.227....1K}.

As implied by Eq. (\ref{eq:Tb}), the fluctuations of differential brightness temperature  is thus reflecting the neutral hydrogen density fluctuation, 
\begin{equation}\label{delta21_simple}
    \delta_{21}(z)\equiv\frac{\Delta T_b-\Delta\bar{T}_b}{\Delta\bar{T}_b}\\
    =\delta_{\rm HI}(z)=\frac{\rho_{\rm HI}(z)-\bar{\rho}_{\rm HI}}{\bar{\rho}_{\rm HI}}.
\end{equation}
The power spectrum $P_{21}(\bold{k})$ of 21cm signal is defined by
\begin{equation}
    \left\langle\Tilde{\delta}_{21}(\bold{k}_1)\Tilde{\delta}_{21}(\bold{k}_2)\right\rangle\equiv(2\pi)^3\delta_{\rm D}^{(3)}(\bold{k}_1+\bold{k}_2)P_{21}(\bold{k}_1),
\end{equation}
where $\Tilde{\delta}_{21}(\bold{k})$ is the Fourier transform of $\delta_{21}(\bold r)$, and $\delta_{\rm D}^{(3)}$ is the Dirac delta function. 

\subsection{Perturbation Theory and Biasing Model}

Calculating a functional form of the 21cm power spectrum requires an expression for the 21cm brightness fluctuation. As shown in Eq.~(\ref{delta21_simple}), this fluctuation is related to fluctuations in the neutral hydrogen density field and can therefore be theoretically predicted using cosmological perturbation theory. The H\,{\sc i} density is a tracer of the matter density field, and in the perturbation theory framework is related to it via a biasing model. This generally takes the form
\begin{equation}\label{eq:biasing}
    \delta_t(\mathbf{r})=\sum_Ob_OO(\mathbf{r}),
\end{equation}
where we have used the subscript $t$ to denote an arbitrary tracer. Here, $O$ are statistical fields that may impact the tracer's density during its formation and the corresponding bias parameter $b_O$ is a number that is usually extracted from simulations. Each $O$ is related to the matter density contrast $\delta$. Note that the basic idea of perturbation theory is that the expansion terms in Eq.~(\ref{eq:biasing}) could extend to arbitrarily high orders.

As in \citet{2020PhRvD.102b3515G}, we take the biasing terms that are needed to compute the leading-order streaming velocity correction to the power spectrum, and that are allowed by symmetry considerations. This includes all second-order gravitational terms\footnote{Third-order terms contribute to the 1-loop power spectrum, but if they do not contain the streaming velocity they will not be part of the streaming velocity correction.}, as well as terms through third order that contain the streaming velocity $v_s$:
\begin{equation}\label{eq:delta21}
 \begin{aligned}
    \delta_{21}(\mathbf{s}) &= c_{1}\delta(\mathbf{r})+c_{2}s_{zz}(\mathbf{r})+c_{3}\delta^{2}(\mathbf{r})+c_{4}s^{2}(\mathbf{r})+c_{5}\delta(\mathbf{r})s_{zz}(\mathbf{r})\\ &+c_{6}t_{zz}(\mathbf{r})+c_{7}s_{zz}^{2}(\mathbf{r})+c_{8}[s_{xz}^{2}(\mathbf{r})+s_{yz}^{2}(\mathbf{r})]\\ &+c_{0} + b_{v}[v_{s}^{2}(\mathbf{x})-1] +b_{vz}\left[v_{s,z}^{2}(\mathbf{x})-\frac{1}{3}v_{s}^{2}(\mathbf{x})\right] + \cdots,
 \end{aligned}
\end{equation}
where the coordinates $\mathbf{s}, \mathbf{r},\textrm{and } \mathbf{x}$ denote redshift space, real space Eulerian position, and real space Lagrangian position, respectively. In this expansion, $\delta$ is the matter overdensity field, $s_{ij}$ is the tidal tensor, $s^2 = s_{ij}s_{ij}$, $t_{ij}$ is a tensor related to the matter overdensity and velocity divergence, $v_s^2$ is the isotropic streaming velocity field, $v_{s,z}^2$ is the quadrupolar streaming velocity field, and $c_0$ is a counterterm chosen to ensure $\langle \delta_{21}(\mathbf{s}) \rangle=0$.

Moreover, \citet{2020PhRvD.102b3515G} provides a straightforward pipeline to calculate perturbative corrections to the power spectrum given a set of bias parameters. We therefore adopt their formalism to calculate the H\,{\sc i} power spectrum. In Section \ref{sec:methodology}, we present the formalism for calculating input bias coefficients and show how to extract them from hydrodynamics simulations.

\subsection{Cosmology}

Throughout this work, we use cosmological parameters from the Planck 2015 ``TT+TE+EE+lowP+lensing+ext'' \citep{2016A&A...594A..13P}: $\Omega_m=0.3089$, $\Omega_{\Lambda}=0.6911$, $\Omega_bh^2=0.02230$, $H_0=67.74\,{\rm km\,s^{-1}\,Mpc^{-1}}$, $Y_p=0.249$, $\sigma_8=0.8159$ and $n_s=0.9667$.

\section{Methodology}
\label{sec:methodology}
\begin{figure*}
    \centering
    \includegraphics[width=\textwidth]{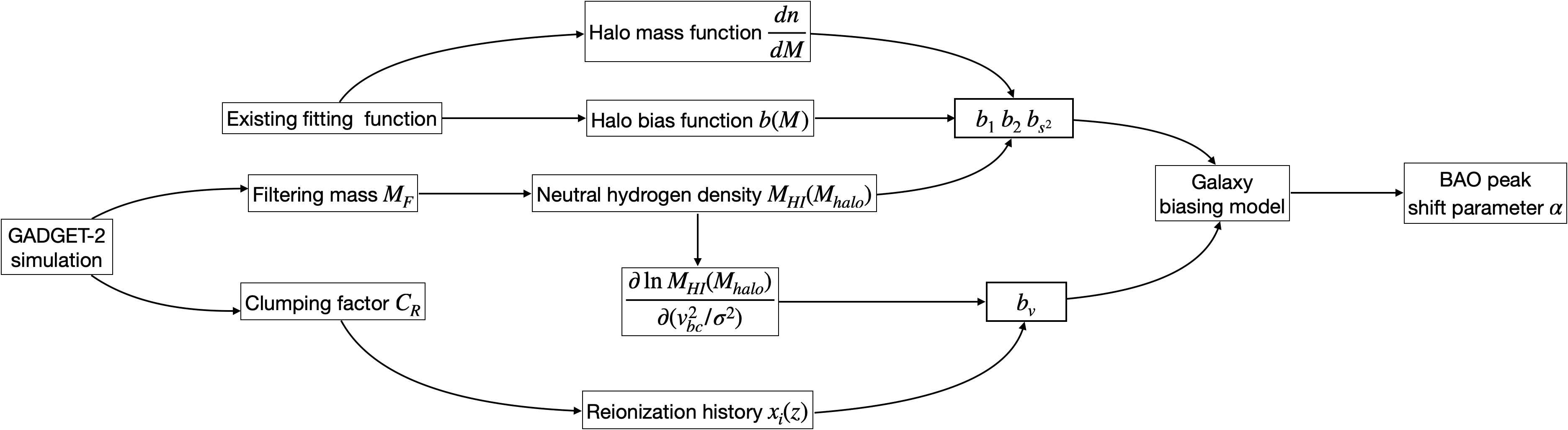}
    \caption{Flow chart from simulation to calculation of BAO scale shift parameter for this work.}
    \label{fig:flowchart}
\end{figure*}

The post-reionization H\,{\sc i} power spectrum could retain  memory of the streaming velocity between dark matter and baryonic matter which existed before recombination. This is possible since streaming velocity affects the H\,{\sc i} distribution in two ways. First, the pre-reionization baryonic structure in mini-halos with mass $\lesssim 10^7 M_{\bigodot}$ is directly suppressed by streaming velocity. 
Once reionization destroys these mini-halos, their contents are fed back into the intergalactic medium (IGM). Second, streaming velocity could modulate reionization history via patchy reionization driven by photons within mini-halos, thereby indirectly modulating the post-reionization matter distribution. Therefore, when constructing our biasing model to calculate the post-reionization H\,{\sc i} power spectrum, we should take these two factors into account in the calculation of the streaming velocity biasing coefficient $b_v$\footnote{These factors are also important for calculating $b_{vz}$. However, $|b_{vz}|$ is sufficiently small compared to $|b_v|$ that we ignore it in our analysis.}. 
As shown in Figure \ref{fig:flowchart}, our work is based on a Gadget-2 simulation of matter evolution. We start by calculating filtering mass $M_F$, which can quantify H\,{\sc i} distribution within halos, to characterize direct effect of streaming velocity. We later account for the indirect effect in the calculation of reionization history. Next, we derive bias coefficients $b_v, b_1, b_2, b_{s^2}$ and use them in our H\,{\sc i} power spectrum calculation. Finally, we fit the power spectrum to obtain the BAO peak shift $\Delta \alpha$ due to streaming velocity.

In the rest of this section, we describe our model for the calculation of power spectrum in \S \ref{subsec:ps}. We show how to account for direct and indirect streaming velocity effect respectively in \S\S\ref{subsec:dir_bv} and \S\S\ref{subsec:indir_bv} and then present the equation to obtain the total streaming velocity bias parameter $b_v$ in \S\S\ref{subsec:bv}. In \S\S \ref{subsec:bias} we present the method of calculating other bias parameters $b_1, b_2, b_{s^2}$. In \S\S \ref{subsec:simulation}, we lay out our Gadget-2 hydrodynamics simulations.

\label{subsec:ps}
To calculate the H\,{\sc i} auto-power spectrum accounting for streaming velocity contributions in redshift space, we use the power spectrum model presented in Eq.~(45) of \citet{2020PhRvD.102b3515G}. The inputs for the power spectrum calculation are transfer functions, linear matter power spectra, and bias coefficients ${ b_1, b_2, b_{s^2}}$ and $ b_v$. We use CLASS \citep{2011JCAP...07..034B} to calculate the transfer functions and linear matter power spectrum. The mapping between galaxy bias parameters $\{ b_1, b_2, b_{s^2} \}$ and generalized coefficients $\{c_1 \ldots c_8\}$ in Eq.~(\ref{eq:delta21}) is shown in \citet[Table II]{2020PhRvD.102b3515G}.

\subsection{Direct Effect of Streaming Velocity}
\label{subsec:dir_bv}
The direct effect on the post-reionization IGM is characterized by filtering masses, halos with masses below the filtering mass rarely hold baryonic matter. The abundance of neutral hydrogen with and without streaming velocity should be different since they have different filtering masses. The post-reionization neutral hydrogen density $\rho_{HI}$ is calculated by
\begin{equation}\label{eq:rho_HI}
    \rho_{\rm HI}(z_{\rm re},z_{\rm obs})=\int dM_{\rm halo}\frac{dn(M_{\rm halo},z)}{dM_{\rm halo}}M_{\rm HI}(M_{\rm halo},z_{\rm obs}, z_{\rm re})
\end{equation}
where $dn/dM$ is the halo mass function \citep{2008ApJ...688..709T}, $M_{\rm HI}$ is the H\,{\sc{i}} mass within a halo of mass $M_{\rm halo}$ at redshift $z$. 

Next, we need a model for the H\,{\sc i}-halo mass relation, $M_{\rm HI}(M_{\rm halo},z)$. This mapping is not directly constrained at high redshift. However, there are a few possible ingredients in this model. The most basic ingredient -- and the one that is central to the physics in this paper -- is that low-mass halos are unable to accrete photoionized gas from the intergalactic medium. We follow \citet{2000ApJ...542..535G} in setting this cutoff at the filtering mass $M_F$
\begin{equation}\label{eq:baryonic_mass}
    M_{\rm HI}(M_{\rm halo},z_{\rm obs},z_{\rm re})\propto \frac{f_b M_{\rm halo}}{[1+(2^{1/3}-1)M_F(z)/M_{\rm halo}]^3}
\end{equation}
\begin{equation}
    M_{\rm HI}(M_{\rm halo},z_{obs}|\,z_{re}) = \int_{z_{\rm re}}^{z_{\rm obs}} M_{\rm HI}(M_{\rm halo},z_{\rm obs},z_{\rm re})\frac{dx_i}{dz_{\rm re}}dz_{\rm re}
\end{equation}
where $f_b\equiv\Omega_b/\Omega_m\approx0.1573$ is the universal baryon fraction. (If all accreted baryons were in the form of H\,{\sc i}, the constant of proportionality would be unity, but for the results of this paper only proportionalities matter.) The filtering mass is related to filtering scale as $M_F=\frac43\pi\rho_m(\frac{\pi}{k_F})^3$. We present the derivation of analytical expression for filtering scale $k_F$ in Appendix \ref{app:filtering}.

In the real Universe, the accreted hydrogen is distributed among several different phases in the interstellar medium and circumgalactic medium of the host, and thus the $M_{\rm HI}(M_{\rm halo})$ relation could differ quite substantially from a strict proportionality with a low-mass cutoff. Normally this is described with a power law index $\alpha$: $M_{\rm HI}\propto M_{\rm halo}^\alpha$. At the high mass end, and at low redshift, group and cluster halos have less H\,{\sc i} than a simple proportionality predicts; the phenomenological high-mass cutoff of \citet{2010MNRAS.407..567B} was used in some early intensity mapping studies, and fits to simulations including AGN feedback give $\alpha<1$ at high masses \citep{2016MNRAS.456.3553V}. Additionally, feedback mechanisms could impose a minimum mass greater than the filtering mass $M_F$. 
In general, we expect the impact of streaming velocities to be enhanced if $\alpha<1$ (since more of the H\,{\sc i} is in halos near the filtering mass), but suppressed if (for example) supernova feedback creates an $M_{\rm min}$ larger than the filtering mass.

One way to assess the importance of these effects is via hydrodynamic simulations. The $M_{\rm HI}(M_{\rm halo})$ mapping inferred from the IllustrisTNG simulations at $z=4$ gives $M_{\rm HI}/M_{\rm halo}$ peaking at 0.027 at $M_{\rm halo}\approx 10^{11} h^{-1}M_\odot$, decreasing to 0.007 if we go down to $M_{\rm halo}= 10^{9} h^{-1}M_\odot$ \citep[Table 1]{2018ApJ...866..135V}, a factor of 4 fall-off. However, as seen in Figure 4 of \citet{2018ApJ...866..135V}, the scatter in the $M_{\rm HI}(M_{\rm halo})$ relation is large, especially at low halo masses; for our purposes, we want the arithmetic average $\langle M_{\rm HI}\rangle(M_{\rm halo})$. This is shown in Figure 7 of \citet{2018ApJ...866..135V}, and $\langle M_{\rm HI}\rangle/M_{\rm halo}$ is seen to vary by no more than a factor of 2 (peak-to-valley) from the highest masses all the way down to the filtering mass cutoff at both $z=4$ and $z=5$. (We have explicitly checked this using the halo catalogs from \citealt{2018ApJ...866..135V}; for example, at $z=4$, $\langle M_{\rm HI}\rangle/M_{\rm halo}$ varies from a maximum of 0.32 down to 0.16 at $M_{\rm halo}=5\times 10^8\,M_\odot/h$, just above the filtering mass.) Given this result, we have not chosen to implement a correction to the simple scaling in Eq.~(\ref{eq:baryonic_mass}).

While total H\,{\sc i} is not directly measured at high redshifts, the bias of damped Lyman-$\alpha$ (DLA) absorbers can provide a constraint on the model, since almost all H\,{\sc i} is found in DLAs. \citet{2018MNRAS.473.3019P} find a DLA bias of $1.92\pm 0.20$ in their highest-redshift bin, $2.5<z<3.5$. This is consistent with the H\,{\sc i} bias $b_1$ that we will infer from Eq.~(\ref{eq:baryonic_mass}), although one should keep in mind that the linear bias of the DLAs is a single number and so could be consistent with a range of models with other values of the power-law index $\alpha$ and cutoff masses \citep[e.g.][]{2017MNRAS.471.1788C}.

\subsection{Indirect Effect of Streaming Velocity}
\label{subsec:indir_bv}

The indirect effect of streaming velocity works by modulating the local reionization history, which is traced by the ionized fraction of hydrogen with respect to redshift, $x_i(z)$. We follow the approach in \citet{2020ApJ...898..149D} and \cite{2020ApJ...898..168C} to calculate the reionization history.

The ``accounting equation'' below considers two competing processes during reionization history: the ionization of neutral hydrogen by emitted photons and the local recombination of H\,{\sc{ii}}
\begin{equation}\label{eq:xi}
    \Dot{x_i}=\frac{\epsilon}{\left< n_H\right>}-\left<C_R\right>\alpha_Bn_ex_i.
\end{equation}
Here, $x_i$ is the ionized fraction, $\epsilon$ is the proper ionizing emissivity (the number of ionizing photons per unit time, per unit volume, produced by the sources) \citep{2015ApJ...802L..19R}, $\left< n_{\rm H}\right>$ is the mean proper hydrogen number density, $n_e$ is the proper free electron number density and $\alpha_B$ is the case B recombination rate for hydrogen. $\left<C_R\right>$ is given by
\begin{equation}\label{eq:C_R}
    \left<C_R\right>(t)=\int^{z(t)}_{z_0}dz_{re}C_R(z_{\rm re},v_{\rm bc},z)P_{z_{\rm re}}(x_i(z))
\end{equation}
where $z_{re}$ is the reionization redshift, $P_{z_{re}}$is the probability distribution of $z_{re}$.
\begin{equation}
    P_{z_{\rm re}}(x_i(t))=\frac{dx_i/dz_{\rm re}}{x_i(t)-x_i(z_0)}
\end{equation}
and the clumping factor $C_R$ is an indicator of local clumpiness, defined as the ratio of true recombination rate to that in a uniform-density IGM with constant temperature $T_{\rm ref}=10^4 K$
\begin{equation}\label{eq:C_R_sim}
C_R\equiv\frac{\left<\alpha_Bn_en_{\rm HII}\right>}{\alpha_B(T_{\rm ref})\left<n_e\right>\left<n_{\rm HII}\right>}
\end{equation}
where we approximate $\left<n_{\rm HII}\right>\approx n_{\rm H}(z)$ and $\left<n_e\right>\approx n_{\rm H}(z)+n_{\rm He}(z)$, i.e. hydrogen is completely ionized after $z_{re}$ and helium is singly ionized. For case B recombination rates $\alpha_B$, we use the function fitted in \citet{1991A&A...251..680P},
\begin{equation}\label{eq:alpha_B}
    \alpha_B(T)=10^{-13}\frac{4.309\times(10^{-4}T)^{-0.6166}}{1+0.6703\times(10^{-4}T)^{0.53}} {\rm cm^3s^{-1}}.
\end{equation}

\subsection{Bias Coefficient $b_v$}
\label{subsec:bv}
Note that the direct effect is quantified in Eq. (\ref{eq:rho_HI}) by gas density $\rho_{\rm HI}(z_{\rm obs},z_{\rm re})$ within halos. The indirect effect is included in the modulated reionization history $x_i$ calculation by Eq.(\ref{eq:xi}). 
We may combine these two effects in the integral below to get the post-reionization H\,{\sc{i}} density with respect to $z_{obs}$ 
\begin{equation}\label{eq:rho_hi}
    \rho_{\rm HI}(z_{obs})=-\int\rho_{\rm HI}(z_{\rm obs},z_{\rm re})\frac{dx_i}{dz_{\rm re}}dz_{\rm re}.
\end{equation}
From here we can obtain the total streaming velocity (direct + indirect) bias coefficient $b_v$ from the equation
\begin{equation}\label{eq:b_v}
\begin{aligned}
    b_{\rm v}(z_{\rm obs})&=\frac{1}{\overline{\rho}_{\rm HI}(z_{\rm obs})}\frac{\partial\rho_{\rm HI}(z_{\rm obs})}{\partial(v_{\rm bc}^2/\sigma^2)}\\
    & = \frac{\rho_{\rm HI}(z_{\rm obs},{\rm with\, SV})-\rho_{\rm HI}(z_{\rm obs}, {\rm no\,SV})}{\overline{\rho}_{\rm HI}(z_{\rm obs})}
\end{aligned}
\end{equation}
where $\sigma=33$\,km \,s$^{-1}$ is the rms value of streaming velocity.
The bias coefficient $b_{\rm v,dir}$ induced only by direct effect is calculated with the H\,{\sc i} density in Eq. \ref{eq:rho_HI} before the reionization modulation is taken into account
\begin{equation}\label{eq:bv_dir}
    b_{\rm v,dir}= \frac{\rho_{\rm HI}(z_{\rm obs},z_{\rm re},{\rm with\, SV})-\rho_{\rm HI}(z_{\rm obs}, z_{\rm re},{\rm no\,SV})}{\overline{\rho}_{\rm HI}(z_{\rm obs},z_{\rm re})}
\end{equation}
and bias due to indirect effect $b_{\rm v,ind}=b_{\rm v}-b_{\rm v,dir}$ therein.

\subsection{Computation of $b_1, b_2, b_{s^2}$}
\label{subsec:bias}
The bias parameter $b_1$ describes how well the 21~cm brightness temperature fluctuations trace the linear matter density fluctuations. It is modelled as
\begin{equation}
    b_1=\frac{\int\frac{d\overline{n}}{dM}b(M)M_{\rm HI}(M)dM}{\int\frac{d\overline{n}}{dM}M_{\rm HI}(M)dM}
\end{equation}
where $b(M)$ is the halo bias \citep{2010ApJ...724..878T}.
The second halo bias parameter $b_2$ is coupled to the second-order matter density fluctuation $\delta^2-\left<\delta^2\right>$ and is calculated by \citep{2012PhRvD..86h3540B}
\begin{equation}
\begin{aligned}
    b_2(M;z)=&\frac{8}{21}\left[\frac{a\nu^2(M;z)-1}{\delta_c}+\frac{2p}{\delta_c{1+[a\nu^2(M;z)]^p}}\right]\\ &+\frac{\nu^2(M;z)-3}{\sigma^2(M;z)}+\frac{2p}{\delta_c^2{1+[a\nu^2(M;z)]^p}}\\
    &\times[2p+2a\nu^2(M;z)-1]
\end{aligned}
\end{equation}
where $\nu=\delta_c/\sigma(M)$, the critical collapsing linear overdensity is $\delta_c=1.686$, and $\sigma$ is the root-mean-square variance of the linear density field smoothed over a filtering scale. Here we take values $p=0.3$ and $a=0.707$ for Sheth-Tormen halo mass functions \citep{1999MNRAS.308..119S}. 
Another bias parameter we need is $b_{s^2}$ which is coupled to the square of the tidal field $s^2$ and simply related to $b_1$ by \citep{2012PhRvD..86h3540B}
\begin{equation}
    b_{s^2}=-\frac{2}{7}(b_1-1).
\end{equation}

\subsection{Simulations and Extraction of Quantities}
\label{subsec:simulation}

We use a modified version of GADGET-2 \citep{2001NewA....6...79S, 2005Natur.435..629S} which was used previously in \citet{2018MNRAS.474.2173H} for our simulations. All simulations start at the time of recombination, $z_{\rm dec}=1059$ with modified initial condition generators to enable or disable streaming velocity between baryons and dark matter. Reionization is implemented by resetting the temperature of gas particles to $2\times 10^4$ instantaneously at $z_{\rm re}$. Each simulation has the same box size, $L=1152{\rm\,h^{-1}\,kpc}$, and the same number of particles, $N=2\times(256)^3$. This is the same mass resolution that was tested and used in \citet{2018MNRAS.474.2173H}. All of our simulations were run on the Ruby and Pitzer clusters at the Ohio Supercomputer Center \citep{Ruby2015}. We archived our modified Gadget-2 N-GenIC files, analysis tools and tabulated filtering mass data in a Github repository\footnote{URL: https://github.com/CosmoSheep/HIPowerSpectrum}.

To calculate the direct effect of streaming velocity, we run simulations with $z_{re} \in \{6,7,8,8.5,9,10,11,12\}$,  $v_{bc}\in \{0,33\}$~km/s, and $z_{obs}$ from 5.5 to 3.5 for each combination. We simulate four realizations to reduce the statistical error due to the limited box size by a factor of $\sqrt{4}$. In the calculation of filtering mass, we need sound speed $c_s$ (see \S\S \ref{ss:thermal}), expressed as 
\begin{equation}
    c_s=\sqrt{\frac{\gamma k_BT_0}{\mu}}
\end{equation}
where $\gamma$ is the index in $T=T_0\Delta^{\gamma-1}$, $\Delta=\rho/\overline{\rho}$ is the overdensity, $k_B$ is Boltzmann's constant, $T_0$ is the mean gas temperature, $\mu$ is the reduced mass at pure hydrogen plasma.

By extracting $c_s$ from simulations and then substituting it into filtering scale calculation, we could get filtering mass and $\rho_{\rm HI}(z_{\rm obs},z_{\rm re})$. We show filtering masses with discrete $z_{\rm obs}$ in Table \ref{table:mf}. The continuous plots for filtering masses are shown in Figure \ref{fig:mf_plot}.

To obtain reionization history we run simulations with $z_{re} \in \{6,7,8,8.5,9,10,11,12\}$ and $v_{bc}\in \{0,33\}$~km/s. We extract $C_R$ from simulations with a cut-off matter density $\rho<300\overline{\rho}$, such that ultra-dense regions that could self-shield from being ionized will not be counted in the process of reionization. To carry out the integral in Eq. (\ref{eq:C_R}), we do 2-dimensional interpolations to obtain continuous $C_R(z_{\rm re},v_{\rm bc},z_{\rm obs})$ in the range of $8\le z_{\rm re}\le12$ and $0\le z_{\rm re}-z_{\rm obs}\le6.0$. Note that we interpolate $C_R$ in Eq.~(\ref{eq:C_R_sim}) in \{$z_{re},z_{re}-z$\} instead of \{$z_{re},z$\} because this way the domain of validity ($z_{re}-z>0$) is aligned with the coordinate axes.
Also, it is reasonable to interpolate through the time after reionization $z_{re}-z_{obs}$ since the changes of $C_R$ for different $z_{re}$ are qualitatively consistent after $z_{re}$, as shown in Figure \ref{fig:CR}.

\section{Results}
\label{sec:results}

\subsection{Filtering Masses}

\begin{figure*}
\centering
\begin{varwidth}{\textwidth}
    \vspace{0pt}
    \includegraphics[width=0.48\textwidth]{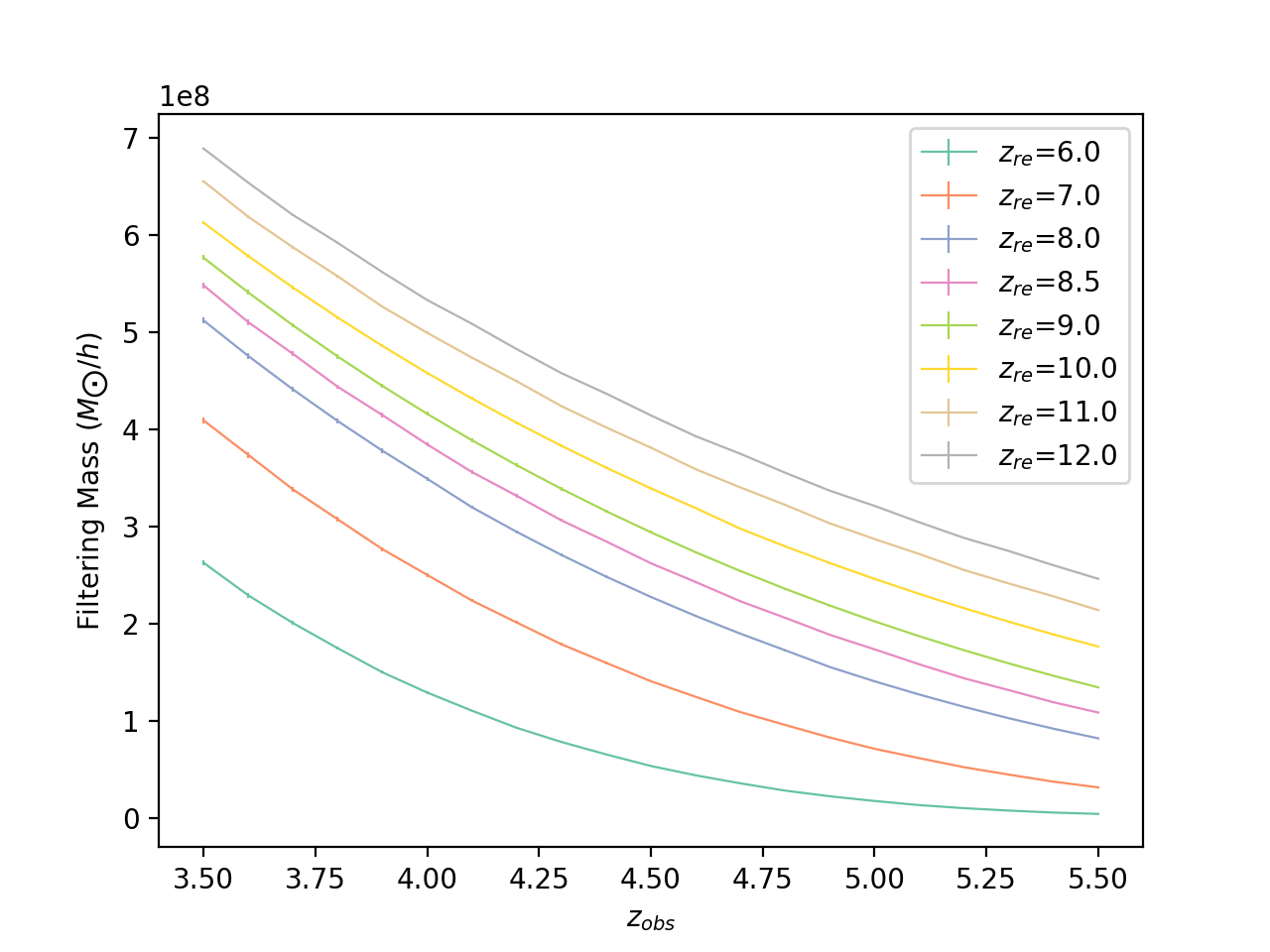}
\end{varwidth}
\begin{varwidth}{\textwidth}
    \vspace{0pt}
    \includegraphics[width=0.48\textwidth]{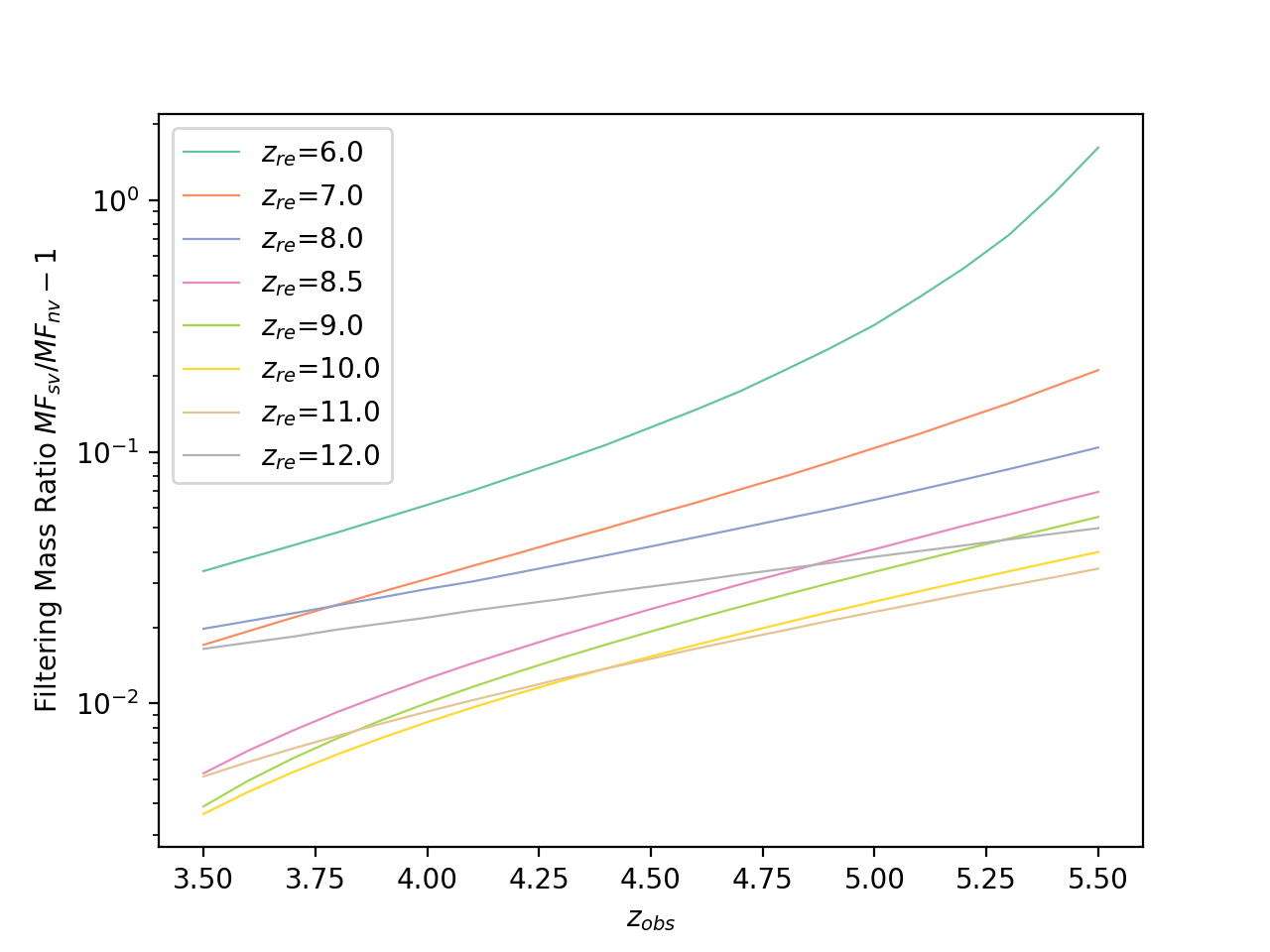}
\end{varwidth}
\caption{Left panel: filtering mass vs. observational redshift $z_{obs}$. Right panel: the ratio of filtering masses with and without streaming velocity vs. $z_{obs}$.}
\label{fig:mf_plot} 
\end{figure*}

We show the filtering masses over the range $6\le z_{\rm re}\le 12$ and $3.5\le z_{\rm obs}\le 5.5$ in Table. \ref{table:mf}. Streaming velocity suppresses the formation of small-scale structures prior to reionization and leads to larger filtering masses compared to results without streaming velocity. We see that the effect of streaming velocity on filtering mass gradually becomes negligible after reionization while the gas has time to relax.  Furthermoer, in both cases filtering masses decrease as $z_{\rm re}$ decreases, since it takes time for the pressure increase at reionization to smooth out small-scale baryonic structure and thereby raise the filtering mass. So after reionization, the filtering masses grow over time. The quantitative properties of filtering masses $M_F$ with respect to $z_{re}$ and $z_{ob}$ are also shown in the left panel of Figure~\ref{fig:mf_plot}. We plot the log of ratio $M_F$ with and without streaming velocity in the right panel of Figure~\ref{fig:mf_plot}. 
\begin{table*}
\caption{Filtering masses in  units of $10^7M_{\bigodot}/h$.}
\begin{tabular}{|c|c|c|c|c|c|c|}
\hline
\hline
$z_{re}$ & $v_{bc}$ & $z_{obs}=5.5$ & $z_{obs}=5.0$ & $z_{obs}=4.5$ & $z_{obs}=4.0$ & $z_{obs}=3.5$\\ \hline
6.0       & on        &$0.384\pm0.003$        &$1.715\pm0.017$        &$5.319\pm0.054$        &$12.877\pm0.133$        &$26.246\pm0.288$ \\ 
      & off        &$0.147\pm0.002$        &$1.300\pm0.015$        &$4.728\pm0.047$        &$12.130\pm0.118$        &$25.393\pm0.260$ \\ 
\hline
7.0       & on        &$3.116\pm0.021$        &$7.087\pm0.047$        &$14.043\pm0.097$        &$24.990\pm0.188$        &$40.898\pm0.337$ \\ 
      & off        &$2.572\pm0.015$        &$6.421\pm0.034$        &$13.299\pm0.073$        &$24.232\pm0.147$        &$40.210\pm0.277$ \\ 
\hline
8.0       & on        &$8.162\pm0.041$        &$14.036\pm0.074$        &$22.722\pm0.130$        &$34.887\pm0.214$        &$51.186\pm0.330$ \\ 
      & off        &$7.393\pm0.026$        &$13.186\pm0.050$        &$21.804\pm0.095$        &$33.920\pm0.160$        &$50.192\pm0.245$ \\ 
\hline
8.5       & on        &$10.818\pm0.051$        &$17.326\pm0.087$        &$26.185\pm0.139$        &$38.419\pm0.209$        &$54.794\pm0.297$ \\ 
      & off        &$10.117\pm0.031$        &$16.644\pm0.055$        &$25.578\pm0.093$        &$37.943\pm0.147$        &$54.506\pm0.220$ \\ 
\hline
9.0       & on        &$13.407\pm0.062$        &$20.208\pm0.096$        &$29.378\pm0.142$        &$41.565\pm0.198$        &$57.641\pm0.265$ \\ 
      & off        &$12.706\pm0.036$        &$19.555\pm0.059$        &$28.821\pm0.093$        &$41.151\pm0.134$        &$57.417\pm0.186$ \\ 
\hline
10.0       & on        &$17.607\pm0.072$        &$24.577\pm0.095$        &$33.893\pm0.120$        &$45.777\pm0.148$        &$61.243\pm0.182$ \\ 
      & off        &$16.930\pm0.045$        &$23.968\pm0.061$        &$33.380\pm0.078$        &$45.394\pm0.097$        &$61.020\pm0.122$ \\ 
\hline
11.0       & on        &$21.362\pm0.068$        &$28.678\pm0.082$        &$38.083\pm0.097$        &$49.905\pm0.114$        &$65.498\pm0.137$ \\ 
      & off        &$20.653\pm0.042$        &$28.029\pm0.050$        &$37.518\pm0.059$        &$49.446\pm0.070$        &$65.164\pm0.087$ \\ 
\hline
12.0       & on        &$24.582\pm0.054$        &$32.092\pm0.061$        &$41.405\pm0.069$        &$53.282\pm0.080$        &$68.876\pm0.098$ \\ 
      & off        &$23.417\pm0.202$        &$30.908\pm0.240$        &$40.232\pm0.273$        &$52.137\pm0.303$        &$67.758\pm0.340$ \\ 
\hline
\hline
\end{tabular}\label{table:mf}
\end{table*}


\label{subsec:filtering_mass}

\subsection{Reionization History}

\begin{figure}
    \centering
    \includegraphics[width=\columnwidth]{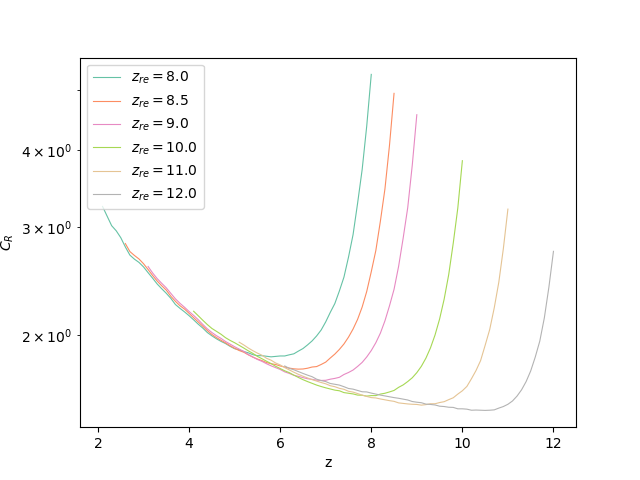}
    \caption{Clumping factors evolution after reionization with $z_{re}=$ 8.0, 8.5, 9.0, 10.0, 11.0, 12.0. The curves drop right after reionization because of the sudden heat of gases by shocks and the tails converge since the gases built pressure equilibrium again after they got time to relax.}
    \label{fig:CR}
\end{figure}

We plot the clumping factor $C_R$ as a function of redshift $z$ with the reionization redshift $z_{re}$ set from 8 to 12 in Figure~\ref{fig:CR}. We see that in each case the clumping factor drops rapidly after reionization because the high-density structures (e.g., filaments and minihalos) are disrupted. The gas in these structures flows out into the lower-density IGM. However, as structure growth continues at larger scales, the clumping factor $C_R$ begins to increase again. The clumping factor curves converge for different reionization redshifts after the relaxation period.

\begin{figure}
    \centering
    \includegraphics[width=\columnwidth]{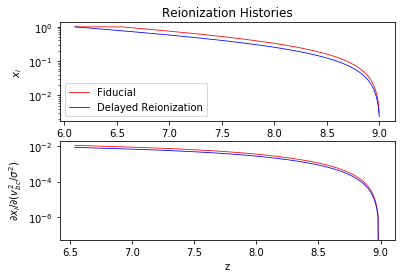}
    \caption{Top panel: The evolution of ionized hydrogen fraction $x_i$ after reionization for $z_{re}=9$. Bottom panel: $x_i$ difference between simulations with and without streaming velocity.}
    \label{fig:rei_his}
\end{figure}

In Figure~\ref{fig:rei_his}, we show the reionization history as the evolution of ionized hydrogen fraction $x_i$ and its derivative with respect to normalized streaming velocity $v^2_{\rm bc}/\sigma^2$. This is calculated by the finite difference between two simulations with $v_{\rm bc}=0$ (off) and $v_{\rm bc}=\sigma=33\,$km\,s$^{-1}$ (the rms value):
\begin{equation}
\frac{\partial x_i}{\partial(v^2_{\rm bc}/\sigma^2)}=x_i(v_{\rm bc}=\sigma, z_{\rm re})-x_i(v_{\rm bc}=0,z_{\rm re}).
\end{equation}
In the $z_{re}=9.0$ fiducial model, reionization is 50\% complete at $z=7.54$, and finishes at $z=6.53$. The Thomson optical depth is $\tau=0.059$, as compared to the {\slshape Planck} measurement of $\tau = 0.054\pm 0.007$ \citep{2020A&A...641A...1P}. We find that the dependence of ionization fraction on streaming velocity $\partial x_i/\partial(v^2_{\rm bc}/\sigma^2)$ is positive (regions of higher streaming velocity have less clumping and reionize faster). The dependence reaches $\partial x_i/\partial(v^2_{\rm bc}/\sigma^2)=0.0092$ at the end of reionization and the reionization ends earlier of $\triangle z =0.02$ due to streaming velocity. We also consider two different reionization scenarios, one delays reionization by dividing the emissivity $\epsilon$ (in Eq. \ref{eq:xi}) by a factor of 1.3, the other starts reionization at $z_{\rm re}=12.0$. We summarize quantites of the 3 reionization history in Table. \ref{table:reion}
We re-compute the indirect term $b_{v,ind}$ using Eqs.~(\ref{eq:rho_hi},\ref{eq:b_v}) for this delayed reionization history; we find that $b_v$ is reduced from $-0.0243$ to $-0.0257$ at $z=5.5$, and $-0.0032$ to $-0.0038$ at $z=3.5$.
\begin{table*}
\centering
\caption{Summary of different reionization histories}\label{table:reion}
\begin{tabular}{c|c|c|c|c|c|c}
\hline
\hline
$z_{re}$ &  & $z_{\rm mid}$ & $z_{\rm end}$ & $z_{\rm end,SV}-z_{\rm end,NV}$ & $\partial x_i/\partial(v^2_{\rm bc}/\sigma^2)$ & $\tau$ \\
\hline
9.0  & Fiducial & 7.54 & 6.53 & 0.02 & 0.0092 & 0.059 \\  
     & Delayed  & 7.20 & 6.09 & 0.02 & 0.0091 & 0.056 \\
\hline
12.0 & Fiducial & 8.35 & 6.82 & 0.01 & 0.0080 & 0.065 \\  

\hline
\hline
\end{tabular}
\end{table*}


\subsection{Bias Parameters}

In Table~\ref{table:bias}, we list the calculation results of bias parameters $b_1$, $b_2$, $b_{s^2}$, and $b_v$ over the range $3.5\le z\le 5.5$. The streaming velocity bias is broken down separately into the direct and indirect effects, $b_{v,dir}$ and $b_{v,ind}$, 
calculated using Eq.\ref{eq:bv_dir}. We see that $b_v<0$ by our calculation, indicating that the streaming velocity reduces the H\,{\sc i} density, and hence that the BAO ruler stretches because of streaming velocity \citep{2016PhRvL.116l1303B}. The absolute value of $b_v$ goes down from $z_{obs}=$5.5 to 3.5, which is consistent with our interpretation of filtering masses in Section~\ref{subsec:filtering_mass}, i.e., the effect of streaming velocity  becomes 
weaker in lower redshifts. Note that the total streaming velocity bias $b_{\rm v}$ is dominated by $b_{\rm v,dir}$ at higher redshifts. But as redshifts go lower, the contribution from $b_{\rm v,ind}$ become more comparable. This indicates that effects of streaming velocity on small-scale structures leave their imprints mainly by directly suppressing the mini-halo abundance, while the memory of ionizing photon sinks modulation is non-negligible when the total streaming velocity memory becomes weaker at lower redshifts.

\begin{table*}
\caption{ Bias parameters for three reionization scenarios. The statistical uncertainties of $b_1$, $b_2$, $b_{s^2}$ are within 1\%. We show the statistical uncertainties of streaming velocity bias parameter from 4 simulation runs in the table.}
\begin{tabular}{|c|c|c|c|c|c|c|c|}
\hline
\hline
$z_{\rm re}$ & $z_{\rm obs}$ & $b_1$ & $b_2$ & $b_{s^2}$ & $b_{\rm v}$ & $b_{\rm v,dir}$ & $b_{\rm v,ind}$\\
\hline
\multirow{5}{2em}{12.0} 

&
3.5
 & $2.2429$ & $3.0579$ & $-0.3551$ & $-0.0041\pm0.0005$ & $-0.0028\pm0.0006$ & $-0.0013\pm0.0001$\\  
&
4.0
 & $2.3722$ & $3.5488$ & $-0.3921$ & $-0.0072\pm0.0006$ & $-0.0049\pm0.0007$ & $-0.0023\pm0.0001$\\  
&
4.5
 & $2.4918$ & $4.0383$ & $-0.4262$ & $-0.0122\pm0.0006$ & $-0.0083\pm0.0008$ & $-0.0038\pm0.0002$\\  
&
5.0
 & $2.5976$ & $4.5004$ & $-0.4565$ & $-0.0195\pm0.0006$ & $-0.0136\pm0.0008$ & $-0.0059\pm0.0002$\\  
&
5.5
 & $2.6888$ & $4.9193$ & $-0.4825$ & $-0.0272\pm0.0005$ & $-0.0213\pm0.0008$ & $-0.0058\pm0.0003$\\  

\hline
\hline
\multirow{5}{2em}{9.0} 
 
&3.5
 & $2.2210$ & $2.9774$ & $-0.3489$ & $-0.0043\pm0.0004$ & $-0.0032\pm0.0007$ & $-0.0010\pm0.0002$\\  

&4.0
 & $2.3374$ & $3.4099$ & $-0.3821$ & $-0.0076\pm0.0005$ & $-0.0057\pm0.0007$ & $-0.0019\pm0.0003$\\  

&4.5
 & $2.4391$ & $3.8139$ & $-0.4112$ & $-0.0130\pm0.0005$ & $-0.0097\pm0.0008$ & $-0.0033\pm0.0003$\\  

&5.0
 & $2.5235$ & $4.1660$ & $-0.4353$ & $-0.0208\pm0.0004$ & $-0.0159\pm0.0008$ & $-0.0049\pm0.0003$\\  

&5.5
 & $2.5928$ & $4.4629$ & $-0.4551$ & $-0.0273\pm0.0004$ & $-0.0243\pm0.0007$ & $-0.0030\pm0.0003$\\ 
 
\hline
\hline
\multirow{5}{2em}{9.0 (Delay)} 
 &
 3.5
 & $2.2063$ & $2.9247$ & $-0.3447$ & $-0.0018\pm0.0003$ & $-0.0038\pm0.0005$ & $0.0020\pm0.0002$\\  

&
4.0
 & $2.3139$ & $3.3200$ & $-0.3754$ & $-0.0054\pm0.0004$ & $-0.0067\pm0.0006$ & $0.0013\pm0.0002$\\  

&
4.5
 & $2.4041$ & $3.6726$ & $-0.4012$ & $-0.0115\pm0.0003$ & $-0.0115\pm0.0006$ & $-0.0000\pm0.0003$\\  

&
5.0
 & $2.4761$ & $3.9662$ & $-0.4217$ & $-0.0199\pm0.0003$ & $-0.0185\pm0.0006$ & $-0.0014\pm0.0003$\\  

&
5.5
 & $2.5403$ & $4.2299$ & $-0.4401$ & $-0.0241\pm0.0003$ & $-0.0257\pm0.0006$ & $0.0017\pm0.0003$\\

\hline
\hline
\end{tabular}\label{table:bias}
\end{table*}

\subsection{BAO Peak Shift}
We determine the BAO scale shift owing to streaming velocity in the following way: We start with an isotropic, no streaming velocity galaxy power spectrum built from the first six terms of Eq. (A7) in \citet{2016PhRvL.116l1303B}. This base power spectrum ($P_{\textrm{base}}$) is fit to a model power spectrum template using $\chi^2$ minimization. The model is
\begin{equation}
    P_{\textrm{model}}(k) = \sum_{j=0}^{2}a_{j}k^{j}P_{\textrm{evo}}
(k/\alpha) + \sum_{j=0}^{5}b_{j}k^{j},
\end{equation}
where $\alpha$ parametrizes the BAO scale and the coefficients $a_{j}$ and $b_{j}$ are marginalized over. Nonlinear BAO damping is factored into this model through an evolved power spectrum 
\begin{equation}
P_{\textrm{evo}}(k)=[P_{\textrm{lin}}(k)-P_{\textrm{nw}}(k)]e^{-k^{2}\Sigma^{2}/2}+P_{\textrm{nw}},
\end{equation}
where $P_{\textrm{lin}}$ is the linear matter power spectrum, $P_{\textrm{nw}}$ is the no-wiggle power spectrum of \citet{1998ApJ...496..605E}, and $\Sigma$ is a damping parameter. The $\chi^2$ integral we minimize is 
\begin{equation}\label{eq:chi2}
    \chi^{2} = V\int_{k_{\textrm{min}}}^{k_{\textrm{max}}}\frac{d^{3}k}{(2\pi)^{3}}\frac{[P_{\textrm{base}}(k)-P_{\textrm{model}}(k)]^{2}}{2[P_{\textrm{base}}(k)+1/\bar{n}]^{2}},
\end{equation}
where $\bar{n}$ is the galaxy number density and $V$ is the volume. These values are given in \citet{2018arXiv181009572C} for the redshifts of interests. The integration range we consider is $0.02\,h\,\textrm{Mpc}^{-1}< k < 0.35\,h\,\textrm{Mpc}^{-1}$. 

Our minimizer uses a Nelder-Mead optimizer to fit the $\chi^2$ integral. It explores parameter space to find $[\,\alpha,\Sigma\,]$ and uses least squares fitting to get associated values of $a_{j}$ and $b_{j}$. We restrict the minimizer to acceptable regions of parameter space by forcing the integral to return a divergent result if it ventures into prohibited regions. We fit for three different $\mu$ values to account for anisotropic damping of the BAO feature. There is no noise in our fits because the matter power spectra are taken from CLASS and processed through FAST-PT \citep{2016JCAP...09..015M,2017JCAP...02..030F}.

Each of the preceding steps are repeated using $P_{21}(k,\mu)$ in place of $P_\textrm{base}(k)$ in Eq. (\ref{eq:chi2}). By taking the best-fit $\alpha$ when streaming velocity is turned off ($P_\textrm{base}$) and subtracting it from the best-fit $\alpha$ when streaming velocity is turned on ($P_{21}$) we get a BAO scale shift $\Delta\alpha$. Values we calculated for $\Delta\alpha$ are given in Table \ref{table:scale_shift}. 

\begin{figure}
    \centering
    \includegraphics[width=\columnwidth]{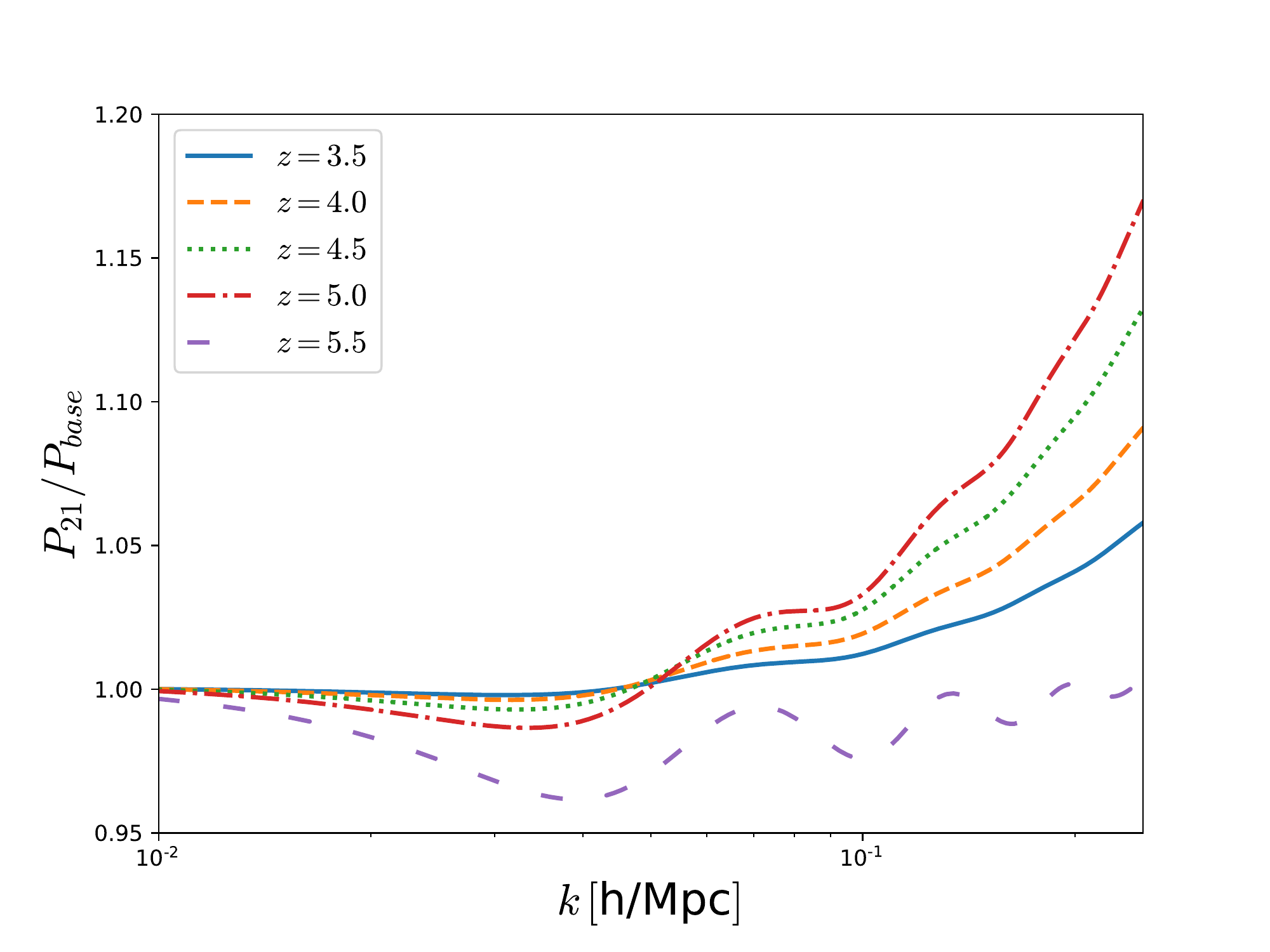}
    \caption{The ratio $P_{21}(k)/P_{\textrm{base}}(k)$ based on bias parameters in Table \ref{table:bias}. This plot displays results for $\mu=1/\sqrt{3}$.}
    \label{fig:power_ratio}
\end{figure}

Figure \ref{fig:power_ratio} shows how much streaming velocity impacts the H{\sc i} power spectrum BAO scale. The  suppression of power in $P_{21}$, seen in the trough near $k=0.04$ $h\,\textrm{Mpc}^{-1}$, is more pronounced at higher redshifts. This is consistent with results in Table \ref{table:scale_shift} showing that streaming velocity effects are larger at higher redshifts.

\begin{table}
\caption{BAO scale shifts as a function of $\mu$ and $z$. We also give the volumes and number densities used in Equation \ref{eq:chi2} since they are functions of redshift.}
\begin{tabular}{lllll}
\hline
\hline
$z_{\textrm{obs}}$ & $V\,[10^{11}\,h^{-3}\textrm{Mpc}^3]$ & $\overline{n}\, [h^3\textrm{Mpc}^{-3}]$ & $\mu$ & $\Delta\alpha$\%\\
\hline
3.5 & 1.8 & 0.030 & 0 & -0.121  \\
 &  &  & $1/\sqrt{3}$ & -0.135  \\
 &  &  & 1 & -0.167  \\
4.0 & 2.2 & 0.033 & 0 & -0.196  \\
 &  &  & $1/\sqrt{3}$ & -0.220  \\
 &  &  & 1 & -0.278  \\
4.5 & 2.5 & 0.033 & 0 & -0.304  \\
 &  &  & $1/\sqrt{3}$ & -0.317  \\
 &  &  & 1 & -0.429  \\
5.0 & 2.8 & 0.031 & 0 & -0.435 \\
 &  &  & $1/\sqrt{3}$ & -0.486 \\
 &  &  & 1 & -0.609 \\
5.5 & 3.1 & 0.028 & 0 & -0.350\\
 &  &  & $1/\sqrt{3}$ & -0.395 \\
 &  &  & 1 & -0.505 \\
\hline
\hline
\end{tabular}\label{table:scale_shift}
\end{table}

\section{Conclusion and Discussion}
\label{sec:discussion}

This work has made a first estimate of the BAO scale shift of post-reionization 21 cm intensity mapping surveys due to the streaming velocity effect. We find that there are two main mechanisms at work. First, there is a ``direct'' effect: the streaming velocity can modulate the amount of pre-reionization small-scale structure, and the destruction of these structures at reionization affects the thermal state and filtering mass of the intergalactic medium. We found that the streaming velocity raises the filtering masses and hence reduces the amount of neutral gas in halos following reionization ($b_{v,\rm direct}<0$).  There is also an ``indirect'' effect since streaming velocities reduce the clumping factor and thus feedback on the local reionization history itself \citep{2020ApJ...898..168C} -- in this case accelerating it, and increasing the post-reionization filtering mass. This effect is minor at higher redshifts but becomes more comparable to the direct effect at lower redshifts while imprints of streaming velocity in total have substantially dissipated.

We predict the bias coefficients at redshifts $3.5 \le z \le 5.5$ and find $b_v<0$, i.e., the BAO scale stretches due to streaming velocity. As one would intuitively expect, $|b_v|$ becomes smaller at later times as the thermal and dynamical memory of reionization is erased. The streaming velocity-induced BAO shifts are $0.167-0.505\%$ in the radial BAO scale and $0.121-0.350\%$ in the transverse BAO scale. These values may be compared against a precision of 0.13\% per $\Delta z=0.5$ bin forecast for the proposed Stage {\sc ii} 21~cm intensity mapping experiment \citep{2018arXiv181009572C}. Although our forecasts are preliminary, this suggests that streaming velocity effects will have to be taken into account in Stage {\sc ii} or a similar future 21 cm intensity mapping experiment. Two other main source of theoretical systematics on BAO scale are nonlinear evolution of the density field and galaxy formation. These two effects could shift BAO peak $0.16\%-0.11\%$ \citep{2009PhRvD..80f3508P} at redshifts 3.5--5.5, which are in the same order of magnitude as streaming velocity effect. Fortunately, these systematics could be substantially reduced by density-field reconstruction as well as further modeling \citep[e.g.][]{2008ApJ...686...13S, 2011ApJ...734...94M}, so they are not expected to be a limiting systematic.

Our estimates in this paper contain several approximations and simplifications that could be relaxed in future work.
We treat photon sinks-modulated reionization history as a local process in our calculation of clumpiness; this is valid as long as the scales of interest ($\sim \pi/k$) are larger than the ionization bubbles, but we expect it to break down toward the later stages of reionization. We leave the non-local modeling of how clumpiness modulated by the streaming velocities affects reionization to future work, since it requires a more elaborate simulation (to capture the range of scales, it would require a large box simulation of reionization with subgrid modeling analogous to \citealt{2006MNRAS.366..689C}, or use of local clumping factors, e.g., \citealt{2007ApJ...657...15K, 2007MNRAS.377.1043M, 2011MNRAS.412L..16R}, based on the small scale clumping factors appropriate to the streaming velocity in that cell). In this work we ignore the streaming velocity effect on the star formation rate, while it could suppress the first stars (Pop III) formation and ionizing photon production and then slow down reionization. \citet{2020ApJ...898..168C} investigates this source bias term, but its impact is still unclear because of large modeling uncertainties. A more accurate quantification of this effect requires future studies with better understanding and modeling of star formation.
We have also ignored X-ray heating prior to reionization, which also reduces the clumpiness of the gas and suppresses small-scale structure. \citet{2018MNRAS.474.2173H} found that extreme models of X-ray heating could reduce $|b_v|$ for the Lyman-$\alpha$ forest, but future studies should check whether this is also true for post-reionization H\,{\sc i}. One should also investigate a wider range of preheating scenarios. Note that we expect some of these potential improvements to the treatment could lower $|b_v|$ (e.g., X-ray heating), some could raise $|b_v|$ (e.g., streaming velocity modulation of Pop III stars), and for some it is not clear what direction to expect (e.g., the non-local treatment of reionization). Thus we interpret our calculation as a reasonable first model, but it is not necessarily an upper or lower bound.

Finally, our results motivate further work on mitigation strategies for the BAO peak shift caused by streaming velocities. Previous works \citep{2015MNRAS.448....9S,2018MNRAS.474.2109S} have shown that 3-point correlation functions could be used to constrain the streaming velocity bias coefficient $b_v$ from galaxy survey data. This approach can also help more accurately measure $b_v$ in future 21 cm intensity mapping experiments and undo the effect of streaming velocity on the BAO scale. Intensity mapping surveys have the subtlety that the mean brightness temperature $\bar T_b(z)$ is not known, which causes a degeneracy for, e.g., growth of structure measurements using redshift space distortions (but see \citealt{2019JCAP...06..025C}), but upon examining the 3-point formulae in \citet{2015MNRAS.448....9S} we do not expect a similar issue will arise for $b_v/b_1$.

A corrected BAO scale will permit us to better constrain the expansion history of the Universe from redshifts $3.5 \leq z \leq 5.5$, and bring us closer to understanding dark energy, including any potential early component.
 
\section*{Acknowledgements}
We thank the anonymous referee, Christopher Cain, Paulo Montero-Camacho, Hee-Jong Seo and Zachary Slepian for useful comments on the draft. 
We thank Francisco Villaescusa-Navarro for making available some of the simulated halo catalogs from \citet{2016MNRAS.456.3553V}.
The authors are supported by NASA grant 15-WFIRST15-0008. This work was partially supported by a grant from the Simons Foundation (\#256298 to Christopher Hirata). JG acknowledges additional support from Princeton's Presidential Postdoctoral Research Fellowship.

This article used resources on the Pitzer Cluster at the Ohio Supercomputing Center.

\section*{Data availability}
The filering mass data and analysis tools underlying this article are archived in a Github repository. All the software used in this manuscript are publicly available. Appropriate links are given in the manuscript.

\appendix

\section{Filtering scale}
\label{app:filtering}

This appendix expresses the filtering scale $k_{\rm F}$ as defined in \citet{1998MNRAS.296...44G} of the baryons as an integral over the thermal history of the Universe. The filtering scale is defined by expanding the ratio of baryonic to dark matter density perturbations in a single Fourier mode as a Taylor series in $k$:  
\begin{equation}
\frac{\delta_b}{\delta_m} \propto 1 - \frac{k^2}{k_{\rm F}^2} + ...\,,
\end{equation}
where only even powers appear because only the magnitude of $k$ matters and the transfer functions for baryons and matter are analytic in $k$. The Green's function solution for $k_{\rm F}^{-2}$ can be completed analytically, in the case of a matter-dominated Universe and assuming instantaneous kinematic decoupling in the early Universe. In an isotropically averaged sense, we can do a similar analytic calculation including the streaming velocities.

This calculation extends the calculation of \citet{2000ApJ...542..535G}, which is equivalent to our result for early decoupling ($a_{\rm dec}\rightarrow 0$), and for which some analytic fitting functions are available \citep{2004ApJ...609..482K}. \citet{2007MNRAS.377..667N} introduced a correction to handle $\delta_b/\delta_m$ not approaching 1 at large scales. There are also extensions for magnetic fields \citep{2008PhRvD..78h3005S, 2010MNRAS.406..482R, 2011MNRAS.410.2149D} and streaming velocities (\citealt{2013ApJ...763...27N}; although not treated the same way as in this paper). One can also find some other calculations in the literature, for example an analytic solution for $\delta_b/\delta_m$ in the case of $c_s^2\propto 1/a$ \citep{2000MNRAS.317..902N}.

The filtering scale has contributions both from gas pressure (the sound speed) and from streaming velocities. We consider the sound speed contribution first.

\subsection{Sound speed}
\label{ss:thermal}

We consider the case of a matter-dominated Universe (so $a\propto t^{2/3}$ and $H=\frac23t^{-1}$). We are interested in the baryon perturbation $\delta_b$ in Fourier mode $k$ at scale factor $a$. The baryon density is taken to be small compared to the matter density. The matter obeys the usual growth function $\delta_m = Ct^{2/3}$. Since only small scales, well below the Silk damping scale, are of interest here, we take an initial condition that the baryons are smoothly distributed at the decoupling epoch, $a_{\rm dec}$, i.e., $\delta_b = \dot\delta_b = 0$. The governing differential equation for baryon perturbations in this case is:
\begin{equation}
\ddot\delta_b + \frac{4}{3t}\dot\delta_b = -\frac{k^2c_s^2}{a^2} \delta_b + \frac{2}{3t^2} \delta_m.
\label{eq:ddot-b}
\end{equation}
In the absence of the sound speed term, this equation has the exact solution
\begin{equation}
\delta_b^{(0)} = C \left(t^{2/3} - 3t_{\rm dec}^{2/3} + 2 t_{\rm dec} t^{-1/3} \right).
\label{eq:db0}
\end{equation}
Here the $t^0$ and $t^{-1/3}$ terms represent the homogeneous solution, with the coefficients chosen to satisfy the initial conditions, and the $t^{2/3}$ term the particular solution.

\begin{figure*}
    \centering
    \includegraphics[width=6.5in]{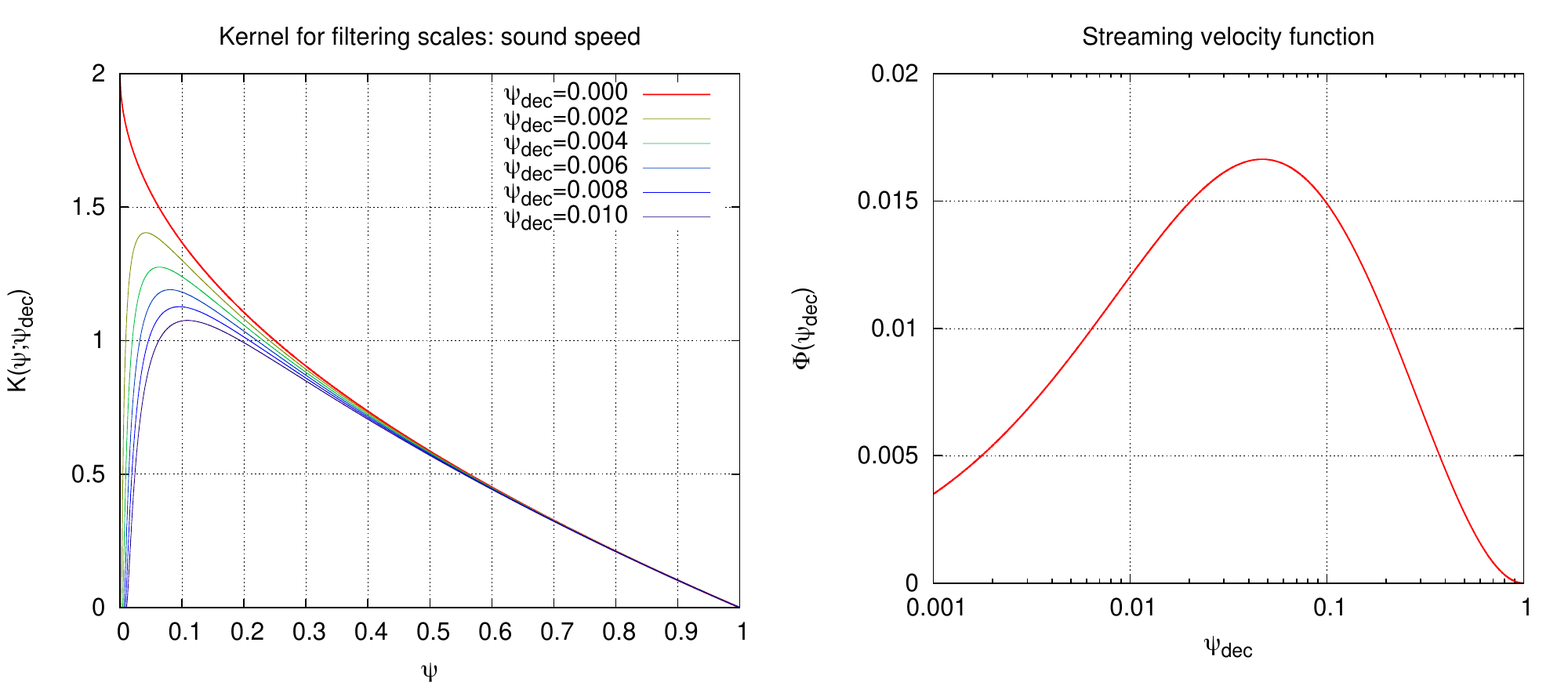}
    \caption{\label{fig:kernel}{\em Left panel}: The kernel $K(\psi;\psi_{\rm dec})$ of Eq.~(\ref{eq:Kernel}), used for the contribution of the sound speed to the filtering length. {\em Right panel}: The function $\Phi(\psi_{\rm dec})$, used to describe the contribution of streaming velocities to the filtering length.}
\end{figure*}

We now treat the sound speed term as a first-order perturbation, i.e., we write $\delta_b = \delta_b^{(0)} + \delta_b^{(1)} + ...$, with the superscript indicating the order in $c_s^2$. To compute $\delta_b^{(1)}$, we use a Green's function approach: we suppose first that $c_s^2/(aH)^2 = \epsilon\delta(t-t_1)$. The superposition of $\delta$ functions can then be used to build up the full solution. The first-order perturbation satisfies
\begin{equation}
\ddot\delta_b^{(1)} + \frac{4}{3t} \dot\delta_b^{(1)} = -\frac4{9t_1^2}k^2\epsilon\delta(t-t_1) \delta_b^{(0)}.
\label{eq:db1}
\end{equation}
The initial condition is that $\delta_b^{(1)}(t_{\rm dec}) = \dot\delta_b^{(1)}(t_{\rm dec}) = 0$; hence $\delta_b^{(1)} = 0$ at $t<t_1$. At $t>t_1$, Eq.~(\ref{eq:db1}) is a homogeneous equation with solutions $\propto t^0$ and $\propto t^{-1/3}$. By requiring continuity of $\delta_b^{(1)}$ at $t_1$, and that the change in $\dot\delta_b^{(1)}$ at $t_1$ is $-\frac49k^2\epsilon \delta_b^{(0)}(t_1)$, we find the coefficients and arrive at
\begin{eqnarray}
\delta_b^{(1)}
&=& \frac43 t_1^{-2/3} t_{\rm dec}^{-1} k^2 \epsilon C \left(t_1^{2/3} - 3t_{\rm dec}^{2/3} + 2 t_{\rm dec} t_1^{-1/3} \right)
\nonumber \\ && \times
\left( -t_{\rm dec}t_1^{-1/3} + t_{\rm dec} t^{-1/3} \right).
\end{eqnarray}
Expanding this as $\delta_b^{(1)} = -(k^2/k_{\rm F}^2) \delta_b^{(0)}$,
and replacing the impulse with an integral over a continuous source $c_s/aH$, we get:
\begin{eqnarray}
k_{\rm F}^{-2} &=& \int_{t_{\rm dec}}^t -\frac43 t_1^{-2/3} t_{\rm dec}^{-1} \frac{ \left(t_1^{2/3} - 3t_{\rm dec}^{2/3} + 2 t_{\rm dec} t_1^{-1/3} \right)  }{ \left(t^{2/3} - 3t_{\rm dec}^{2/3} + 2 t_{\rm dec} t^{-1/3} \right) } 
\nonumber \\ && \times
\left( - t_{\rm dec}t_1^{-1/3} + t_{\rm dec} t^{-1/3} \right)\left.\frac{c_s^2}{(aH)^2}\right|_{t_1} \, dt_1.
\end{eqnarray}
Finally, we may choose to write this in terms of $\psi = a(t_1)/a(t)$, and define $\psi_{\rm dec} = a(t_{\rm dec})/a(t)$. The differential transforms as $t_1 = \psi^{3/2}t$ and $dt_1 = \frac32\psi^{1/2}t\,d\psi$. This leads to
\begin{equation}
k_{\rm F}^{-2} = \int_{\psi_{\rm dec}}^1 K(\psi;\psi_{\rm dec})\,\left.\frac{c_s^2}{(aH)^2}\right|_{t_1}\,d\psi,
\label{eq:kF2}
\end{equation}
where the kernel is
\begin{equation}
K(\psi;\psi_{\rm dec})
                                             = 2  \frac{ [ 1 - 3(\psi_{\rm dec}/\psi) + 2(\psi_{\rm dec}/\psi)^{3/2} ] ( 1 - \psi^{1/2} ) }{ 1 - 3\psi_{\rm dec} + 2 \psi_{\rm dec}^{3/2} } .
\label{eq:Kernel}
\end{equation}
This kernel is shown in the left panel of Fig.~\ref{fig:kernel}.

\subsection{Streaming velocity}

Now we neglect the gas pressure and focus instead on the streaming velocities. The dark matter has a velocity relative to the baryons of $-(t_{\rm dec}/t)^{2/3}{\bf v}_{\rm bc,dec}$, where ${\bf v}_{\rm bc,dec}$ is the streaming velocity at decoupling and we have taken into account the $\propto 1/a$ redshifting of peculiar velocities. This means that there is a comoving displacement between the dark matter and the baryons of
\begin{equation}
\Delta{\xi} = \int -(t_{\rm dec}/t)^{2/3}{\bf v}_{\rm bc,dec}\,\frac{dt}{a}
= 3\frac{{\bf v}_{\rm bc,dec}}{a_{\rm dec}} t_{\rm dec}^{4/3} t^{-1/3}.
\end{equation}
This means that in the reference frame of the baryons, the dark matter perturbation includes a phase shift and grows as $Ct^{2/3}e^{{\rm i}k\mu\Delta{\xi}}$, where $\mu$ is the cosine of the angle between ${\bf k}$ and ${\bf v}_{\rm bc}$. Thus the baryon growth equation is
\begin{eqnarray}
\ddot\delta_b + \frac{4}{3t}\dot\delta_b &=& \frac{2}{3} Ct^{-4/3} e^{{\rm i} g (t_{\rm dec}/t)^{1/3}}
\nonumber \\
&=& \sum_{j=0}^\infty \frac{2}{3} \,\frac{{\rm i}^j}{j!}\, t_{\rm dec}^{j/3} Ct^{-(4+j)/3} g^j,
\label{eq:ddot-b-vbc}
\end{eqnarray}
where $g = 3k\mu v_{\rm bc,dec} t_{\rm dec}/a_{\rm dec}$, and in the second equality we have expanded the exponential in a Taylor series. Each term on the right-hand side is a power law $\propto t^{-(4+j)/3}$, and hence the equation can be satisfied by a power law $\propto t^{(2-j)/3}$ (except for the $j=2$ term, which can be satisfied by $\ln t$). Thus the particular solution through the $j=2$ order contains terms proportional to $t^{2/3}$, $t^{1/3}$, and $\ln t$. We also include in our solution homogeneous solutions proportional to $t^{-1/3}$ and $t^0$ to ensure that the initial condition $\delta_b=\dot\delta_b=0$ at $t_{\rm dec}$ is satisfied:
\begin{eqnarray}
\delta_b &=& C\Bigl[ t^{2/3}-t_{\rm dec}^{2/3} + 3{\rm i}gt_{\rm dec}^{1/3}(t^{1/3}-t_{\rm dec}^{1/3}) - g^2t_{\rm dec}^{2/3}\ln \frac{t}{t_{\rm dec}}
\nonumber \\ && ~
+ (2+3{\rm i}g-3g^2) t_{\rm dec}(t^{-1/3} - t_{\rm dec}^{-1/3})
\Bigr] + {\cal O}(g^3).
\nonumber \\ &&
\end{eqnarray}
Taking the absolute value gives -- after some simplification --
\begin{eqnarray}
|\delta_b| &=& C\Bigl[
t^{2/3} - 3t_{\rm dec}^{2/3} + 2t_{\rm dec}t^{-1/3}
+ g^2 \Bigl(- t_{\rm dec}^{2/3}\ln\frac{t}{t_{\rm dec}}
\nonumber \\ && ~
-3t_{\rm dec}t^{-1/3} + 3t_{\rm dec}^{2/3} + \frac{9t_{\rm dec}^{2/3}(t^{1/3}-t_{\rm dec}^{1/3})^2}{2t^{1/3}(t^{1/3}+2t_{\rm dec}^{1/3})}
 \Bigr)
\Bigr]
\nonumber \\ && ~
+ {\cal O}(g^3).
\end{eqnarray}
Setting this equal to $1-k^2/k_{\rm F}^2 + ...$, and taking the angular average so $\mu^2 \rightarrow\frac13$, we find
\begin{equation}
k_{\rm F}^{-2} = 3\left(\frac{v_{\rm bc,dec}t_{\rm dec}}{a_{\rm dec}}\right)^2 \Phi(\psi_{\rm dec}),
\end{equation}
where
\begin{equation}
\Phi(\psi_{\rm dec}) =
\frac{ 3\psi_{\rm dec}^{3/2} - 3\psi_{\rm dec} -\frac32 \psi_{\rm dec} \ln \psi_{\rm dec}
- \frac{9\psi_{\rm dec}(1-\psi_{\rm dec}^{1/2})^2}{2(1+2\psi_{\rm dec}^{1/2})}
}{ 1 - 3\psi_{\rm dec} + 2\psi_{\rm dec}^{3/2}
},
 \label{eq:Phi}
\end{equation}
where $\psi_{\rm dec} = (t_{\rm dec}/t)^{2/3} = a_{\rm dec}/a$ as in Appendix \ref{ss:thermal}. Note that we had to keep terms through second order in $g$ (or in ${\bf v}_{\rm bc}$ in order to get a non-trivial result. This makes sense because a scalar or isotropically averaged filtering length must be even in ${\bf v}_{\rm bc}$. The function $\Phi(\psi_{\rm dec})$ is positive (as it should be!) and is shown in the right panel of Fig.~\ref{fig:kernel}.

\subsection{Combination}

If we add the lowest order contributions to $|\delta_b|$ from both the sound speed and the streaming velocity, we find that the filtering scales add in inverse quadrature:
\begin{equation}
k_{\rm F,tot}^{-2} = k_{\rm F,sound}^{-2} + k_{\rm F,vbc}^{-2}.
\end{equation}

\end{document}